\documentclass[fleqn,usenatbib]{mnras}

\usepackage{newtxtext,newtxmath}
\usepackage[T1]{fontenc}

\DeclareRobustCommand{\VAN}[3]{#2}
\let\VANthebibliography\thebibliography
\def\thebibliography{\DeclareRobustCommand{\VAN}[3]{##3}\VANthebibliography}

\usepackage{graphicx}	% Including figure files
\usepackage{amsmath}	% Advanced maths commands
\usepackage{amssymb}	% Extra maths symbols
\usepackage{placeins}
\usepackage{float}
\usepackage{gensymb}

\def\tab{Table~}
\def\fig{Fig.~}
\def\g133{G133.50+9.01}

\def\sun{$_\odot$}
\def\um{$\rm \mu m$}

\def\h2{H$_2$}

\title[G133.50+9.01: Cloud-cloud collision candidate]{G133.50+9.01: A likely cloud-cloud collision complex triggering the formation of filaments, cores and a stellar cluster}

\author[Issac et al.]{Namitha Issac$^1$\thanks{E-mail: namithaissac.16@res.iist.ac.in (NI)}, Anandmayee Tej$^1$\thanks{E-mail: tej@iist.ac.in (AT)}, Tie Liu$^{2,3}$, Yuefang Wu$^4 $\\
$^1$Indian Institute of Space Science and Technology, Thiruvananthapuram 695 547, Kerala, India\\
$^2$Shanghai Astronomical Observatory, Chinese Academy of Sciences, 80 Nandan Road, Shanghai 200030, People's Republic of China \\
$^3$Key Laboratory for Research in Galaxies and Cosmology, Chinese Academy of Sciences, 80 Nandan Road, Shanghai 200030, People's Republic of China \\
$^4$Department of Astronomy, Peking University, 100871, Beijing, China\\
}

\date{Accepted XXX. Received YYY; in original form ZZZ}

\pubyear{2020}

\begin{document}
\label{firstpage}
\pagerange{\pageref{firstpage}--\pageref{lastpage}}
\maketitle

\begin{abstract}
We present compelling observational evidence of {\g133} being a bona fide cloud-cloud collision candidate with signatures of induced filament, core, and cluster formation.   
The CO molecular line observations reveal that the {\g133} complex is made of two colliding molecular clouds with systemic velocities, $\rm -16.9\,km\,s^{-1}$ and $\rm -14.1\,km\,s^{-1}$. The intersection of the clouds is characterised by broad bridging features characteristic of collision. The morphology of the shocked layer at the interaction front resembles an arc like structure with enhanced excitation temperature and {\h2} column density.
A complex network of filaments is detected in the SCUBA 850\,{\um} image with 14 embedded dense cores, all well correlated spatially with the shocked layer. A stellar cluster revealed through an over-density of identified Class I and II young stellar objects is found located along the arc in the intersection region corroborating with a likely collision induced origin.   
\end{abstract}


%%%%%%%%%%%%%%%%%%%%%%%%%%%%%%%%%%%%%%%%%%%%%%%%%%

%%%%%%%%%%%%%%%%% BODY OF PAPER %%%%%%%%%%%%%%%%%%

\section{Introduction}

There is growing evidence that cloud-cloud collision can trigger star and cluster formation \citep{{2015ApJ...806....7T},{2017ApJ...835L..14G},{2018PASJ...70S..43S},Liu2018a,Liu2018b}. Hydrodynamic simulations \citep{{1992PASJ...44..203H},{2010MNRAS.405.1431A},{2014ApJ...792...63T}} reveal that collisionally compressed interfaces form dense, self-gravitating cores. Magneto-hydrodynamic (MHD) simulations by  \citet{2013ApJ...774L..31I} and other recent studies \citep[e.g.][]{{2015ApJ...811...56W},{2016ApJ...820...26F},{2018MNRAS.473.4220L},{2017ApJ...835..142T},{2018PASJ...70S..45O},{2019A&A...632A.115G},{2019ApJ...886...15T},{2019arXiv190808404S}} give insightful discussions indicating the importance of cloud-cloud collision in high-mass star and cluster formation. The proposed picture involves supersonic collisions that boost the magnetic field strength and gas density in the shock-compressed layer. This results in an enhanced mass accretion rate and also a large effective Jeans mass, the two prime ingredients for massive star formation. The shocked layer collapses into a filament with an enhanced self-gravity in the post-shock gas. A number of pre-stellar cores, accreting mass from the natal cloud are formed in these filaments. Recent studies on a dark cloud L1158 by \citet{{2017ApJ...835L..14G},{2019A&A...632A.115G}} show evidence of collision triggered cluster formation of low- and intermediate-mass stars as well. Notwithstanding the complexities involved in deciphering cloud kinematics from stellar feedback, observationally, many cloud-cloud collision candidates have been discovered \citep[see][and references therein]{hayashi2020triggered}. If cloud collision is frequent in disk galaxies like Milky Way \citep{2009ApJ...700..358T}, then detailed investigation of cloud-cloud collision candidates can shed crucial light on triggered star formation under different physical conditions. A recent paper by \citet{2020arXiv200905077F} provides an excellent review on the current status of observations and theoretical understanding of cloud-cloud collisions and also discusses future directions in this area.

\par In this paper, we investigate the Planck Galactic Cold Clump (PGCC), G133.50+9.01, using molecular line observation. Located in the field of the classical Cepheid SU Cas, a young cluster associated with G133.50+9.01 is identified by \citet{2012MNRAS.421.1040M} using 2MASS and WISE photometry. The cluster that contains countless young stellar objects (YSOs) deviates from spherical symmetry and exhibits an apparent diameter of 3$\times$6\,arcmin. \citet{2018ApJS..236...49Z} detected 18 dust cores from the 850\,{\um} map. We present new results on this complex based on a comprehensive analysis of the CO line kinematics and morphology in conjunction with infrared and sub-mm data to understand the related stellar population and dust component.  

\section{Observations}

\subsection{PMO observations}

The $\rm ^{12}CO\,(1-0)$ and $\rm ^{13}CO\,(1-0)$ lines were observed with the Purple Mountain Observatory 13.7-m telescope (PMO-13.7 m) in August 2013, as a part of PMO survey of Planck Galactic Cold Clumps \citep{Wu2012,Liu2012,{2018ApJS..236...49Z}}. The nine beam array receiver system in single-sideband mode was used as the front end \citep{shan2012}. The half-power beam width (HPBW) is 56\,arcsec and the main-beam efficiency is $\sim 0.45$. $\rm ^{12}CO\,(1-0)$ and $\rm ^{13}CO\,(1-0)$
lines were obtained simultaneously. FFTS spectrometers were used as back ends, which have a total bandwidth of 1\,GHz and 16384 channels, corresponding to a velocity resolution of $\rm 0.16\,km\,s^{-1}$ for the $\rm ^{12}CO\,(1-0)$ and 0.17 km~s$^{-1}$ for the $\rm ^{13}CO\,(1-0)$. The on-the-fly (OTF) observing mode was utilized. The scan speed is $\rm 20\,arcsec\,s^{-1}$ and the mapping area is 22\,arcmin $\times$ 22\,arcmin. These data are smoothed with a beam size of $\rm \sim 63\,arcsec$. The typical rms noise level per channel is 0.3\,K in T$_A^*$ for the $\rm ^{12}CO\,(1-0)$ and 0.2\,K for the $\rm ^{13}CO\,(1-0)$. 

\subsection{SCUBA 850\,{\um} data}

The dust continuum observation at 850\,{\um} was carried out in November 2014 using Submillimeter Common-User Bolometer Array 2 (SCUBA-2, \citealt{2013MNRAS.430.2513H}) on the James Clerk Maxwell Telescope (JCMT). JCMT is a 15-m telescope single-dish telescope that operates in the submillimetre wavelength region of the spectrum. This instrument can provide an effective beam size of 14.6\,arcsec in the 850\,{\um} band. The ``CV Daisy'' mapping mode was used with a mapping area of about $\rm 12\,arcsec \times 12\,arcsec$. The data are reduced using SMURF in the STARLINK package. The rms level is around $\rm 18.1\,mJy\,beam^{-1}$.

\subsection{Dust polarization data from Planck}
To determine the orientation of the magnetic field in the region associated with {\g133}, we make use of the {\it Planck} 353\,GHz (850\,{\um}) dust continuum polarization data \citep{2016A&A...594A...1P}. The polarization data including the Stokes {\it I}, {\it Q} and {\it U} maps used are from the {\it Planck} Public Data Release 2 Full Mission Map with PCCS2 Catalog\footnote{\url{https://irsa.ipac.caltech.edu/applications/planck/}} \citep{2016A&A...594A..26P}. These maps have a beam size of $\rm \sim 5\,arcmin$ and a pixel size of $\rm \sim 1\,arcmin$.

\subsection{WISE archival data}
To identify the stellar population associated with {\g133}, we use the mid-infrared (MIR) archival data from Wide-field Infrared Survey Explorer (WISE, \citealt{2010AJ....140.1868W}). WISE mapped the sky at the MIR wavelengths 3.4, 4.6, 12, and 22\,{\um} with angular resolutions of 6.1, 6.4, 6.5, and 12.0\,arcsec, respectively. We retrieve sources from the ALLWISE\footnote{\url{http://wise2.ipac.caltech.edu/docs/release/allwise/}} catalog using NASA/IPAC Infrared Science Archive (IRSA). ALLWISE combines the data from the WISE cryogenic and NEOWISE \citep{2011ApJ...731...53M} post-cryogenic survey phases. Combining the two data products, ALLWISE provides enhanced photometric sensitivity and accuracy, and better astrometric precision. 

\section{Results}

\subsection{Distance to {\g133}} \label{distance}
\begin{figure}
\centering 
\includegraphics[scale=0.50]{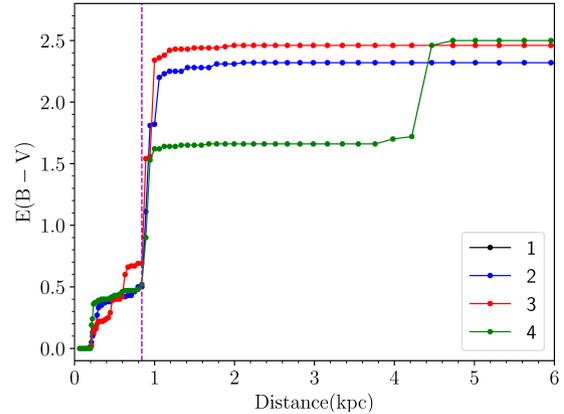}
\caption{The cumulative reddening along four sightlines marked by crosses in {\fig}\ref{CO_mom_spec}(b). The reddening plot towards sightline 1 overlaps with that towards sightline 3. The dashed vertical line indicates the estimated distance to {\g133}.}
\label{distance_plot}
\end{figure}

Citing the nature of CO and HI emission, \citet{2012MNRAS.421.1040M} speculate that the cluster may coincide with the complex in the foreground of SU Cas (which is at a distance of  $\rm 418\pm 12\,pc$)  or, beyond it at a distance, $d \leq \rm 950\,pc$. In their study, \citet{2018ApJS..236...49Z} use the Bayesian distance estimation method proposed by \citet{2016ApJ...823...77R}. Here, trigonometric parallaxes from the Bar and Spiral Structure Legacy Survey and Japanese VLBI Exploration of Radio Astrometry are combined with the probability density function (PDF) for kinematic distance, displacement from the plane, and proximity to individual parallax sources to generate a combined PDF.
Taking the observed centroid velocity of the $\rm ^{13}CO\,(1-0)$ line, \citet{2018ApJS..236...49Z} estimate the distance to {\g133} to be $\rm 0.61\pm0.14\,kpc$.
In this paper, we implement an alternate approach to estimate the distance following the method outlined in \citet{2017ApJ...835L..14G}. This procedure makes use of the 3D dust reddening map\footnote{\url{http://argonaut.skymaps.info}} from \citet{2019ApJ...887...93G}. The dust reddening maps presented by these authors are based on 2MASS and Pan-STARRS~1 photometric data in combination with GAIA parallaxes. 
The distance to estimated by measuring the cumulative reddening, $E(B-V)$, along the line-of-sight. Four sightlines, marked by crosses in {\fig}\ref{CO_mom_spec}(b), are selected. The median cumulative reddening towards these four sightlines are plotted as a function of the distance in {\fig}\ref{distance_plot}. The reddening plot towards the sightlines 1 and 3 overlap. The plot shows a steep rise in $E(B-V)$ at a distance of $\sim \rm 840\,pc$. Such a steep rise can be attributed to the dust reddening in {\g133}. $E(B-V)$ towards the sightline 4 shows an additional steep rise at a distance of $\sim \rm 4.2\,kpc$. This possibly indicates the presence of a background, unrelated cloud. In the analysis presented in this paper, we have taken $\rm 840\,pc$ as the distance to {\g133}.

%%%%%%%%%%% distance %%%%%%%%%%%%%
%%%%%%%%%%%%%%%%%%%%%%%%%%%%%%

\subsection{CO morphology of {\g133} complex} \label{morphology}

%%%%%%%%%% moment 0 %%%%%%%%%%
\begin{figure*}
\centering 
\includegraphics[scale=0.40]{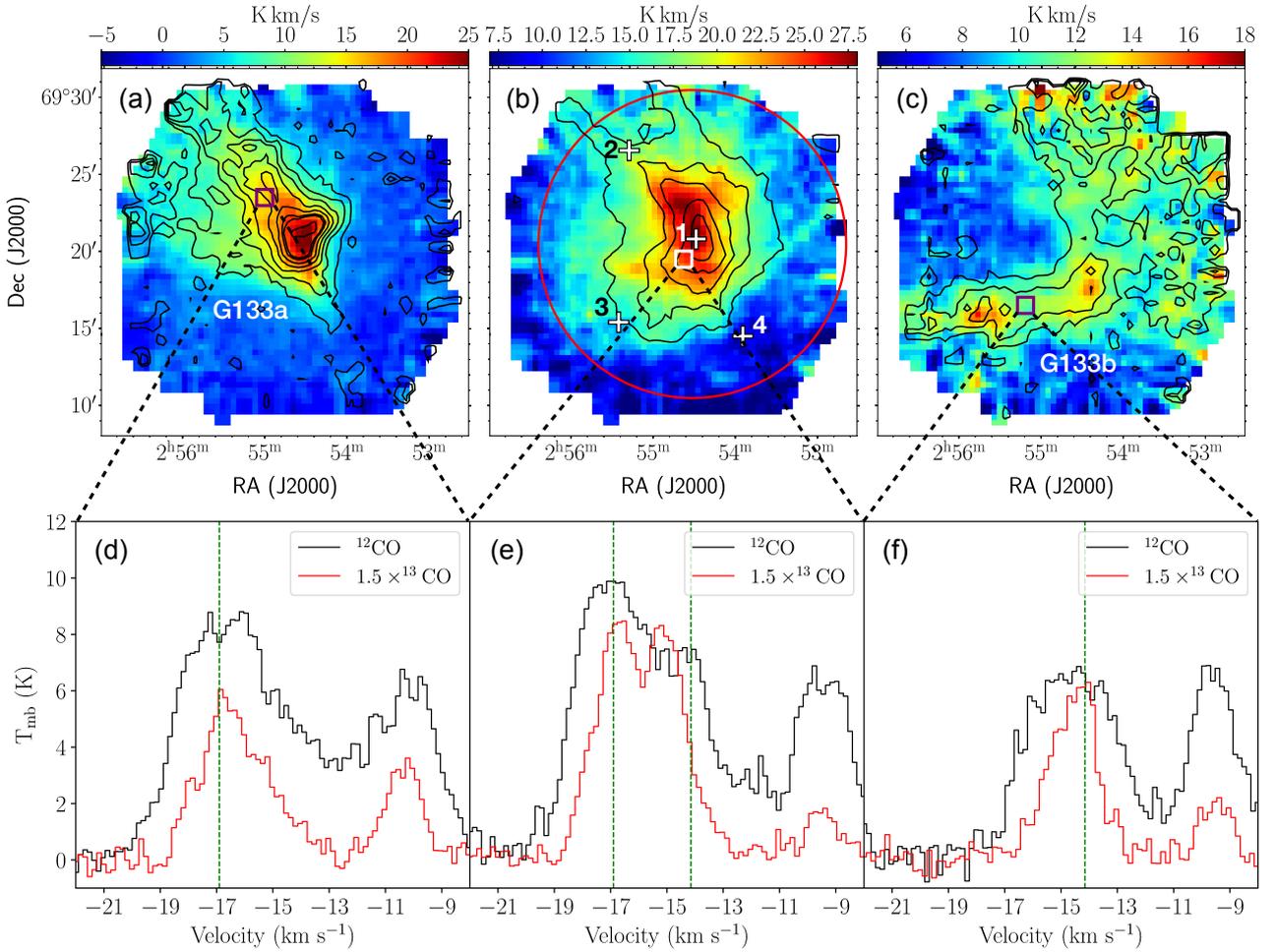}
\caption{(a) The colour scale depicts the $\rm ^{12}CO\,(1-0)$ integrated intensity within the velocity $-19.4$ to $-16.9\,\rm km\,s^{-1}$. The contours of the $\rm ^{13}CO\,(1-0)$ emission over the same velocity range is overlaid in black with levels starting from $3\sigma$ and increasing in steps of $3\sigma$ ($\rm \sigma=0.5\,K\,km\,s^{-1}$). (b) Same as (a) with the integration velocity range $-16.9$ to $-14.1\,\rm km\,s^{-1}$. The $\rm ^{13}CO\,(1-0)$ contours start from $3\sigma$ and increases in steps of $2\sigma$ ($\rm \sigma=1.0\,K\,km\,s^{-1}$). The white crosses mark the regions towards which the median cumulative reddening is estimated to find the distance. The red circle describes a region of radius 10\,arcmin over which the young stellar population is identified (Section~\ref{YSO_population}). (c) Same as (a) with the integration velocity range $-14.1$ to $-11.6\,\rm km\,s^{-1}$. The $\rm ^{13}CO\,(1-0)$ contours start from $3\sigma$ and increases in steps of $2\sigma$ ($\rm \sigma=0.6\,K\,km\,s^{-1}$). (d)-(f) The $\rm ^{12}CO\,(1-0)$ (black) and $\rm ^{13}CO~(1-0)$ (red) spectra extracted at three different positions of the cloud complex {\g133}, indicated by squares in (a)-(c). The vertical green lines indicate the systemic velocities of G133a and G133b as estimated from the $\rm ^{13}CO\,(1-0)$ line}.
\label{CO_mom_spec}
\end{figure*}
%%%%%%%%%%%%%%%%%%%%%%%%%%%%
Morphology of the {\g133} complex is deciphered using molecular line observations of CO. The $\rm ^{12}CO\,(1-0)$ emission is usually considered to be optically thick throughout the molecular cloud, and hence an excellent tracer of its spatial extent, rather than the density distribution. The $\rm ^{13}CO\,(1-0)$ emission, on the other hand, is less optically thick and probes the denser regions of the molecular cloud. 
The integrated intensity maps of $\rm ^{12}CO\,(1-0)$ and $\rm ^{13}CO~(1-0)$ line emission towards {\g133} in three velocity ranges are shown in {\fig}\ref{CO_mom_spec}(a)-(c).  Similar morphology is displayed in both emission lines. The molecular cloud has an elongated morphology in the north-east direction in the velocity range $-19.4$ to $-16.9\,\rm km\,s^{-1}$ as is evident from {\fig}\ref{CO_mom_spec}(a), an arc like morphology in the velocity range $-16.9$ to $-14.1\,\rm km\,s^{-1}$, shown in {\fig}\ref{CO_mom_spec}(b). In the range  $-14.1$ to $-11.6\,\rm km\,s^{-1}$, a hemispherical, `boomerang-like' morphology with a cavity opening towards the north-east and extended emission towards the north-west is noticeable in {\fig}\ref{CO_mom_spec}(c). The emission morphology seen in the above velocity ranges suggests the likely presence of two coalescing molecular clouds (hereafter G133a and G133b) associated with {\g133}. 
The $\rm ^{12}CO\,(1-0)$ and $\rm ^{13}CO~(1-0)$ spectra extracted, sampling these clouds and the intersection region is plotted in {\fig}\ref{CO_mom_spec}(d)-(f). The spectra corresponding to G133a and G133b have single peaked line profiles. From the peak of the $\rm ^{13}CO~(1-0)$ line, the systemic velocities of G133a and G133b are estimated to be $-16.9\,\rm km\,s^{-1}$ and $-14.1\,\rm km\,s^{-1}$, respectively. The intersection region clearly shows two velocity components. 
From the spectra of the interaction region presented, it is seen that for cloud G133b, the peak velocity of the $\rm ^{13}CO~(1-0)$ line is offset from the estimated systemic velocity of $-14.1\,\rm km\,s^{-1}$. This can be attributed to the interaction of the colliding clouds resulting in an intermediate velocity in the intersection region.
 
%%%%%%%%%% channel map %%%%%%%%%%
\begin{figure*}
\centering 
\includegraphics[scale=0.47]{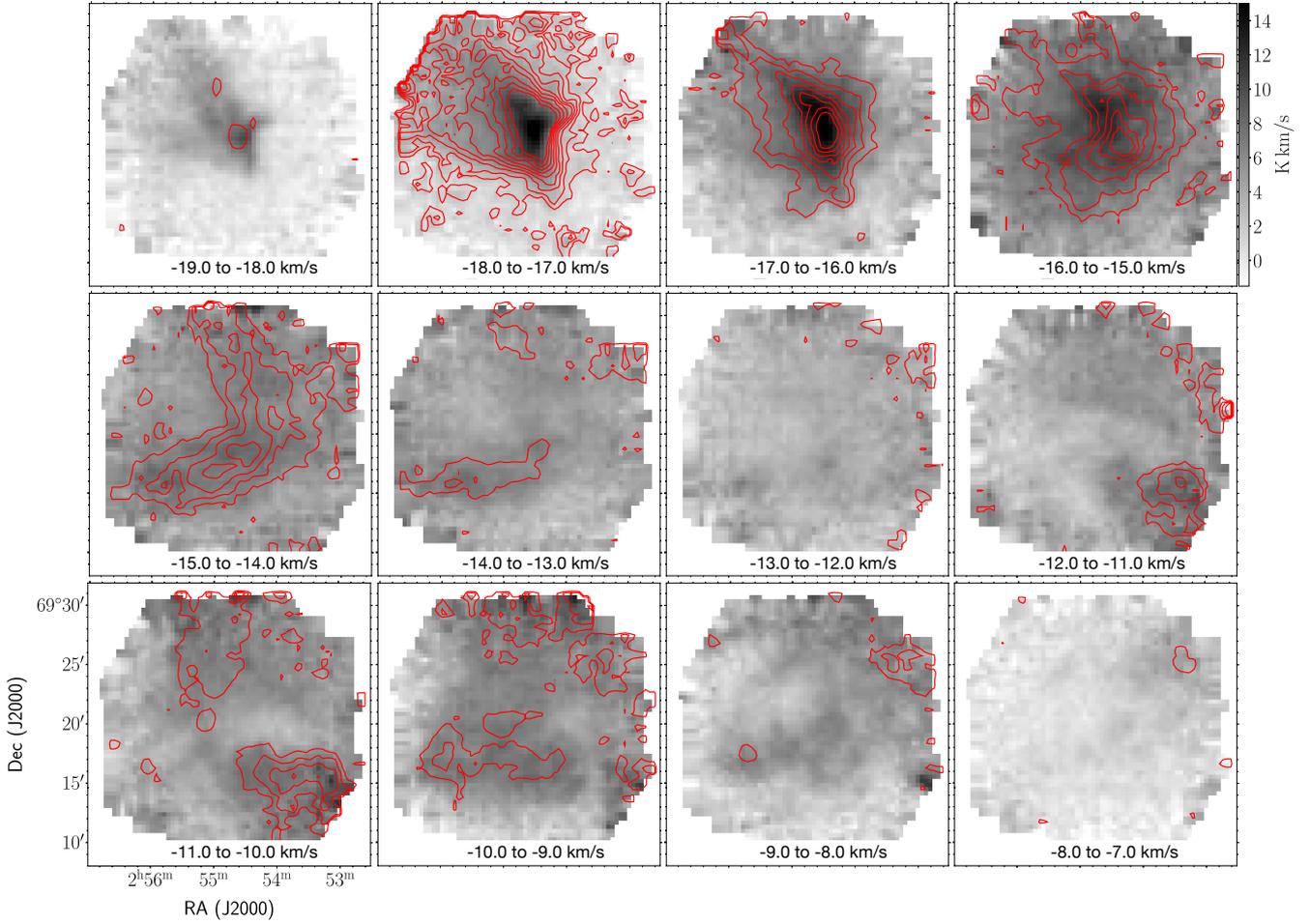}
\caption{Chanel maps for region associated with {\g133}. The grey scale shows the $\rm ^{12}CO\,(1-0)$ channel map. Each map has a channel width of $\rm 1\,km\,s^{-1}$. The contours of the $\rm ^{13}CO\,(1-0)$ emission at each velocity channel is overlaid in red. The contours start at $3\sigma$ ($\rm \sigma=0.5\,K\,km\,s^{-1}$) and increases in steps of $2\sigma$.} 
\label{channel_map}
\end{figure*}
%%%%%%%%%%%%%%%%%%%%%%%%%%%%

\par {\fig}\ref{CO_mom_spec}(d)-(f) show an additional velocity component at $\rm \sim -10\,km\,s^{-1}$ in all three spectra presented. To understand this, we construct channel maps of  $\rm ^{12}CO\,(1-0)$ and $\rm ^{13}CO\,(1-0)$ emission in the velocity range $-19.0$ to $-7.0\,\rm km\,s^{-1}$. These are plotted in {\fig}\ref{channel_map}. As can be seen from the maps, emission from G133a and G133b peaks in the velocity range $-18.0$ to $-16.0\,\rm km\,s^{-1}$ and $-15.0$ to $-13.0\,\rm km\,s^{-1}$, respectively. An arc like morphology, as seen in {\fig}\ref{CO_mom_spec}(b) is evident at intermediate velocities. In the channel maps for the velocity range, $-12.0$ to $-9.0\,\rm km\,s^{-1}$, another component is visible which is consistent with the spectra in {\fig}\ref{CO_mom_spec}(d)-(f). As discussed in Section~\ref{distance}, the reddening plot along the sightline covering this component (\# 4) shows two steep rises at distances 0.84\,kpc and 4.2\,kpc. Given that the velocity is $\rm \sim -10\,km\,s^{-1}$, this cloud is likely to be at a similar distance as that of G133a and G133b with a background, unassociated cloud farther away.

\par Under the assumption of local thermodynamic equilibrium, the total mass of the molecular clouds can be estimated from the optically thin $\rm ^{13}CO$ emission. Following \citet{2016MNRAS.460...82S}, the column density of $\rm ^{13}CO$ is calculated and from the abundance ratio of $\rm ^{13}CO$ to {\h2}, the {\h2} column density is estimated. To determine the $\rm ^{13}CO$ column density, the effective line-of-sight excitation temperature is determined. Combining the radiative transfer equation and the Rayleigh-Jeans law, the brightness temperature can be expressed as,

\begin{equation}
T_{\rm mb}=T_0~\Bigg(\frac{1}{{\rm e}^{T_0/T_{\rm ex}}-1}-\frac{1}{{\rm e}^{T_0/T_{\rm bg}}-1}\Bigg)~(1-{\rm e}^{-\tau})
\label{T_mb}
\end{equation}

\begin{figure*}
\centering 
\includegraphics[scale=0.25]{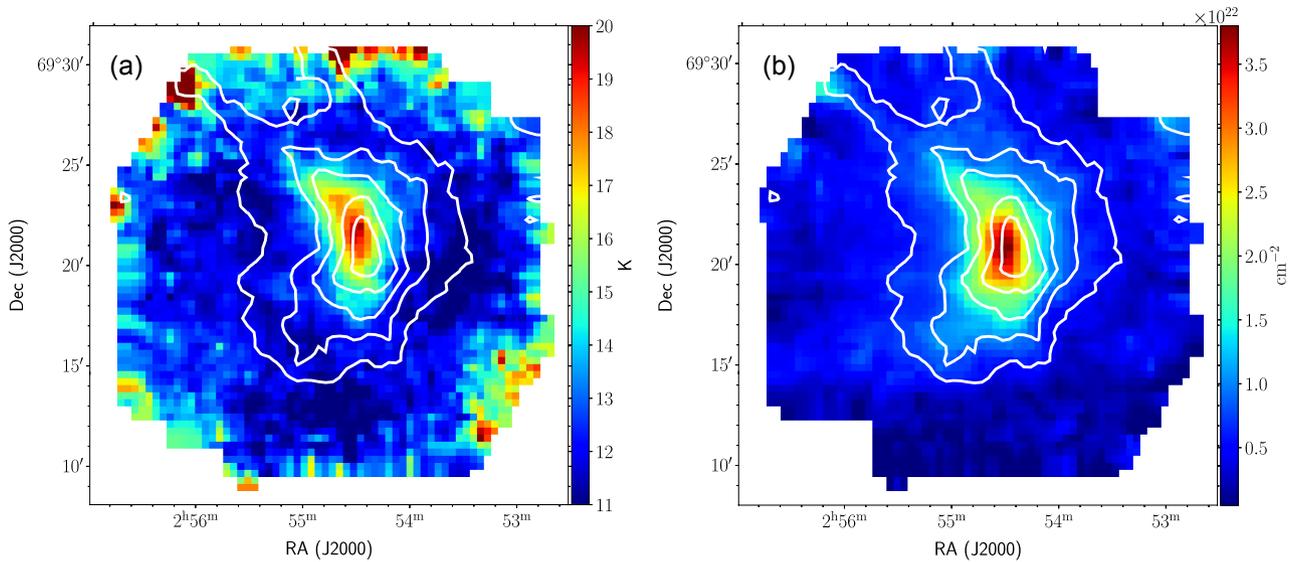}
\caption{(a) The excitation temperature map of {\g133} derived from the peak intensity of the $\rm ^{12}CO\,(1-0)$ emission. (b) The {\h2} column density map of the region associated with {\g133} constructed following the steps described in Section~\ref{morphology}. The white contours correspond to the $\rm ^{13}CO\,(1-0)$ emission within the velocity range $-16.9$ to $-14.1\,\rm km\,s^{-1}$. The contour levels are same as in {\fig}\ref{CO_mom_spec}(b).}
\label{Tex_NH2}
\end{figure*}

\noindent 
Here $T_0=h\nu/k_{\rm B}$, $h$ is the Planck's constant, $k_{\rm B}$ is the Boltzmann's constant, $T_{\rm bg}=2.7\,{\rm K}$ is the background temperature, and $\tau$ is the optical depth.  Assuming that the $\rm ^{12}CO\,(1-0)$ emission is optically thick ($\tau \gg 1$), the excitation temperature per pixel of $\rm ^{12}CO\,(1-0)$ spectral cube is given by

\begin{equation}
T_{\rm ex}=5.5~{\rm ln}~\Bigg(1+\frac{5.5}{T_{\rm mb,peak}^{12}+c_1}\Bigg)^{-1} 
\label{T_ex}
\end{equation}

\noindent
where, $T_{\rm mb,peak}^{12}$ is the peak intensity of $\rm ^{12}CO\,(1-0)$ at each pixel, $5.5\,{\rm K}=h\nu(^{12}{\rm CO})/k_{\rm B}$, and constant $c_1=0.82$ for $T_{\rm bg}=2.7\,{\rm K}$. From the $\rm ^{12}CO\,(1-0)$ spectral cube, the $T_{\rm mb,peak}^{12}$ map is built, from which, using Equation~\ref{T_ex}, the excitation temperature map of $\rm ^{12}CO\,(1-0)$ is constructed which is shown in  {\fig}\ref{Tex_NH2}(a). We assume that the excitation temperatures of $\rm ^{12}CO\,(1-0)$ and $\rm ^{13}CO\,(1-0)$ are equal along the line-of-sight. Substituting for $T_{\rm ex}$ and the $\rm ^{13}CO\,(1-0)$ peak brightness temperature in Equation~\ref{T_mb}, the pixel-wise optical depth can be derived following the equation,

\begin{equation}
\tau_{13}=-{\rm ln}~\Bigg[1-\frac{T_{\rm mb,peak}^{13}}{5.3}\bigg\lbrace \Big({\rm e}^{5.3/T_{\rm ex}}-1\Big)^{-1}-c_2\bigg\rbrace^{-1}\Bigg]
\end{equation}

\noindent
where, $T_{\rm mb,peak}^{13}$ is the peak brightness temperature of $\rm ^{13}CO\,(1-0)$, $5.3\,{\rm K}=h\nu(^{13}{\rm CO})/k_{\rm B}$, and $c_2=0.16$. Assuming that the excitation temperature equals the kinetic temperature for all the energy states and the levels are populated according to Boltzmann distribution, the $\rm ^{13}CO$ column density is calculated from the optical depth using the following expression

\begin{equation}
N({^{13}}{\rm CO})=3.0\times10^{14}\times \frac{\tau_{13}}{1-e^{-\tau_{13}}} \times \frac{\int T_{\rm mb}^{13}(v)dv}{1-e^{-5.3/T_{\rm ex}}}
\end{equation}

\noindent 
$\int T_{\rm mb}^{13}(v)dv$ is the integrated intensity over velocity in units of $\rm km\,s^{-1}$. {\fig}\ref{Tex_NH2}(b) shows the {\h2} column density map constructed with a constant $[\rm ^{12}CO/^{13}CO]$ isotopic ratio of 77 \citep{1994ARA&A..32..191W} and a  $[\rm H_2/^{12}CO]$ abundance ratio of $1.1\times 10^4$ \citep{1982ApJ...262..590F}. 

%%%%%%%%% excitation temperature map %%%%%%%%%
%%%%%%%%%%%%%%%%%%%%%%%%%%%%%%%%%%

\par Considering the region within the $3\sigma$ level of the $\rm ^{13}CO\,(1-0)$ emission integrated over the velocity range $-19.4$ to $\rm -11.6\,km\,s^{-1}$, the mean line-of-sight {\h2} column density, $N({\rm H_2})$, is calculated to be $\rm 8.7\times 10^{21}\,cm^{-2}$. Subsequently, the total mass of the molecular cloud complex ($=N({\rm H_2})\mu_{\rm H_2}m_{\rm H}A$; $A$ is the physical area of the cloud complex, $m_{\rm H}$ is the mass of hydrogen atom) is estimated to be $2.6\times 10^3\,M_\odot$. Here, the mean molecular weight, $\mu_{\rm H_2}$ is taken to be 2.8. The individual masses of G133a and G133b are computed to be $1.1\times 10^3\,M_\odot$ and $1.4\times 10^3\,M_\odot$, respectively, with the assumption that both clouds have equal contribution to the mass at the intersection.

\par {\fig}\ref{Tex_NH2}(a) shows the $\rm ^{12}CO\,(1-0)$ excitation temperature map overlaid with the contours of $\rm ^{13}CO\,(1-0)$ emission integrated over the velocity range $-16.9$ to $-14.1\,\rm km\,s^{-1}$. The intersection of the clouds, G133a and G133b reveals an open arc structure. The $\rm ^{13}CO$ emission also follows a similar arc like morphology. The radius of curvature of the arc is $\rm \sim 1.6\,pc$. The excitation temperatures within this arc varies from $\rm 13-20\,K$ and is higher than that of the ambient cloud ($\rm \sim 11\,K$). The {\h2} column density map given in {\fig}\ref{Tex_NH2}(b) also reveals a density enhancement along this arc. 

\subsection{Cloud kinematics}

%%%%%%%%%%% pv diagram %%%%%%%%%%%%
\begin{figure*}
\centering 
\includegraphics[scale=0.36]{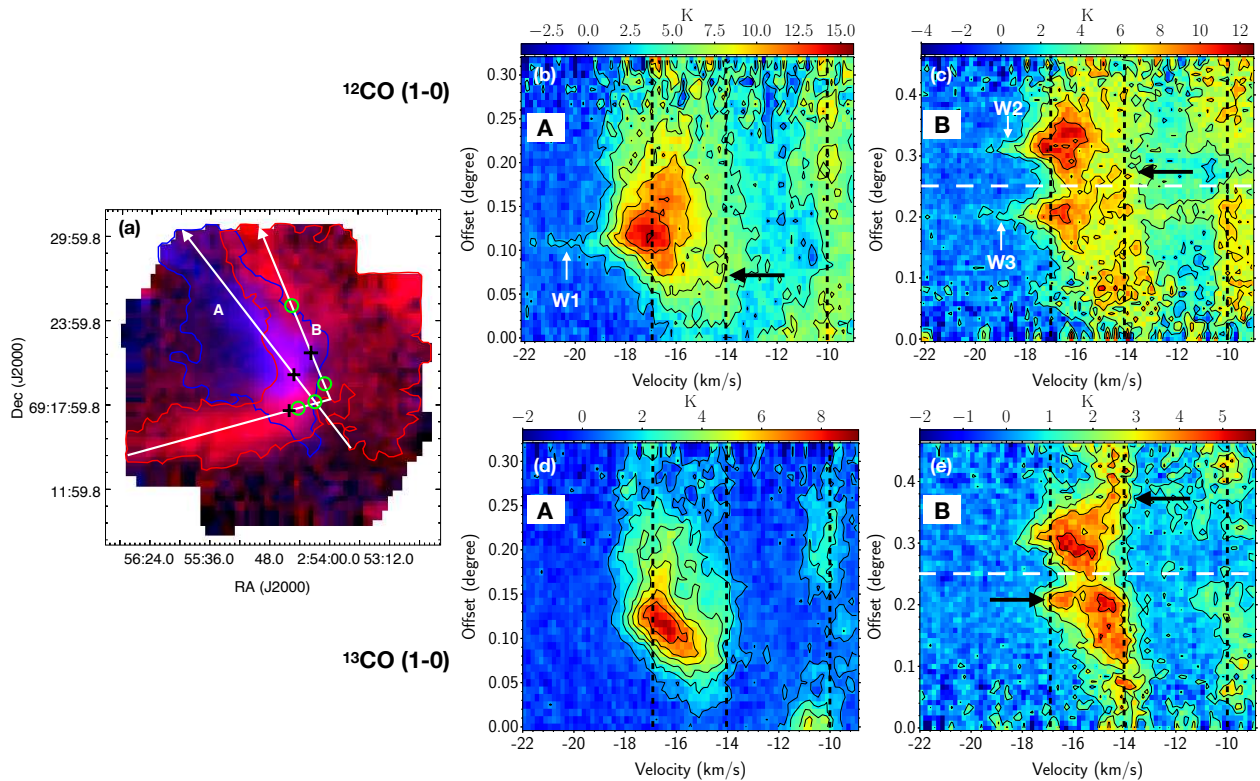}
\caption{(a) Two-colour composite integrate intensity map of $\rm ^{12}CO\,(1-0)$ integrated between $-19.4$ to $-16.9\,\rm km\,s^{-1}$ (blue) and $-14.1$ to $-11.6\,\rm km\,s^{-1}$ (red) of the {\g133} complex. The $3\sigma$ contour of the $\rm ^{13}CO\,(1-0)$ emission (same as in {\fig}\ref{CO_mom_spec}a and c) at both velocity ranges is overlaid in blue and red. A and B are the cuts along which the PV slices are extracted, sampling G133a and G133b, respectively. (b) and (d) PV diagram of $\rm ^{12}CO$ and $\rm ^{13}CO$, respectively along the cut A. (c) and (e) Same as (b) and (d) along the cut B. The horizontal dashed line delineates the PV slices on either side of the bend in B. The contours start at $3\sigma$ and increases in steps of $4\sigma$ ($\rm \sigma = 0.6\,K$ for $\rm ^{12}CO$ and 0.3\,K for $\rm ^{13}CO$). The vertical dashed lines represent the systemic velocities of G133a, G133b and the cloud at $\rm \sim -10\,km\,s^{-1}$. The bridging features with intermediate velocity are marked by black arrows. The positions of the bridging features on the PV cut is marked by green circles in (a). The $\rm ^{12}CO$ wings are indicated by white arrows and their positions on the PV cut are indicated by black crosses in (a).}
\label{pv_diagram}
\end{figure*}
%%%%%%%%%%%%%%%%%%%%%%%%%%%%%%%%

The $\rm ^{12}CO\,(1-0)$ and $\rm ^{13}CO\,(1-0)$ position-velocity (PV) diagram of the {\g133} complex is constructed to understand its velocity structure. {\fig}\ref{pv_diagram}(a) depicts the two-colour composite integrated intensity map of $\rm ^{12}CO\,(1-0)$ integrated between $-19.4$ to $\rm -16.9\,km\,s^{-1}$ (blue) and $-14.1$ to $\rm -11.6\,km\,s^{-1}$(red). The $3\sigma$ contours of the integrated $\rm ^{13}CO\,(1-0)$ within the same velocity ranges is overlaid. PV slices are extracted along two directions, A and B,  highlighted in {\fig}\ref{pv_diagram}(a). These slices probe G133a and G133b, respectively. Both, $\rm ^{12}CO\,(1-0)$ and $\rm ^{13}CO\,(1-0)$ emission show intermediate ``bridging features'' between the peak velocities, $-16.9$ and $\rm -14.1\,km\,s^{-1}$ highlighted by black arrows in {\fig}\ref{pv_diagram}(b), (c) and (e). 
Apart from these, in {\fig}\ref{pv_diagram}(b) and (c), high-velocity $\rm ^{12}CO$ wings are also evident.

\subsection{Dust filaments and cores} \label{dust}

%%%%%%%%%%% 850 %%%%%%%%%%%%
\begin{figure}
\centering 
\includegraphics[scale=0.35]{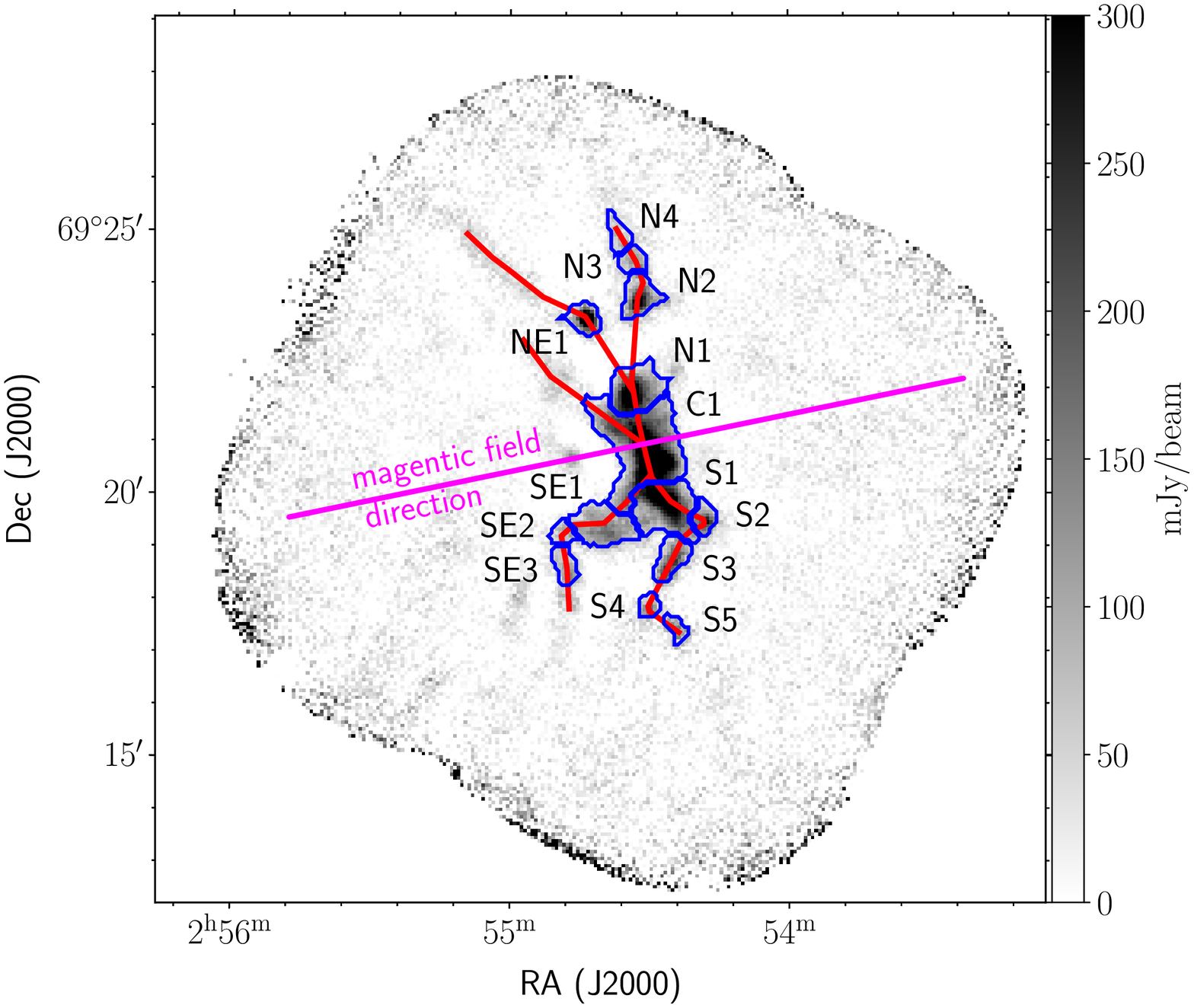}
\caption{The dust emission at 850\,{\um} in the region surrounding {\g133} is presented here. The FellWalker retrieved apertures of the dust cores are outlined in blue and are labelled according to their positions in the field. The skeletons of the visually identified filaments are sketched in red. The magenta line represents the orientation of the magnetic field with respect to the filaments.}
\label{850_clump}
\end{figure}
%%%%%%%%%%%%%%%%%%%%%%%%%%%%%%%%

{\fig}\ref{850_clump} shows the 850\,{\um} dust emission towards {\g133}. A visual inspection of the 850\,{\um} map reveals that the emission from the dust component has a filamentary structure in the north-south direction that bifurcates at the ends. The skeletons of the filaments are overlaid on the image. Localized dust peaks are seen along the filaments. We use the FellWalker clump identification algorithm \citep{BERRY201522}, which is part of the Starlink CUPID package, to identify the dust cores in this region. A detection threshold of $\rm 2.5\sigma~(\sigma=18.1\,mJy\,beam^{-1}$) is used and pixels outside the $2.5\sigma$ level are considered noise. Using this method we detect 14 dust cores along the filaments that are named according to their positions in the field. Relative to the central core, C1, four cores, N1-N4, extend towards the north and five cores, S1-S5, extend towards the south along one of the filaments. The remaining four cores, NE1 and SE1-SE3 lie along the other filaments in the north-east and south-east directions, respectively. The retrieved apertures of these cores are superimposed on the 850\,{\um} map shown in {\fig}\ref{850_clump}. Correlation with {\fig}\ref{Tex_NH2}(b) shows that the detected cores are located within the arc structure displaying enhanced column density. The mass of each core is estimated from the total 850\,{\um} flux density integrated within the core aperture. Assuming thermal emission from optically thin dust, the following expression is used,

\begin{equation}
M_C=\frac{S_\nu d^2}{\kappa_\nu B_\nu (T_d)}
\end{equation}

\noindent
Here, $S_\nu$ is the flux density at 850\,{\um}, $d$ is the distance, $T_d$ is the dust temperature of the cores taken to be the mean excitation temperature, 15.3\,K, within the open arc structure, and $\kappa_\nu=0.1(\nu/1200\,{\rm GHz})^\beta$ is the dust opacity. $\beta$, the dust emissivity index, is assumed to be 2. The effective radii, $r=(A/\pi)^{0.5}$, of the dust cores are determined from the area, $A$, enclosed within the retrieved apertures. Peak positions, radii and masses of the dust cores are tabulated in {\tab}\ref{core_prop}. 

%%%%%%%%%%%%%%%%%%%%%%%%%%%%%%%%%%%%%%%%%%%%%%%%%%%%%%%%%%%%%%%%%%%%%5
\begin{table}
\caption{Physical parameters of the identified dust cores associated with {\g133}. }
\begin{center}

%\hspace*{-1.1cm}
\centering  
\begin{tabular}{l c c c c } \hline \hline 
Dust Core 		&\multicolumn{2}{c}{Peak position} 			& Radius		& Mass			\\

& $\rm \alpha(J2000)~({^h}~{^m}~{^s})$		& $\rm \delta(J2000)~(\degree~\arcmin~\arcsec)$		&(pc)			& (M\sun)	\\

\hline 
\noindent
C1	& 2 54 30.33		& 69 20 39.80 		& 0.20 & 86.0\\
N1	& 2 54 34.11		& 69 21 51.80 		& 0.12 & 27.9\\
N2 	& 2 54 31.84		& 69 23 39.80 		& 0.09 & 10.8 \\
N3 	& 2 54 31.84 		& 69 24 15.80		& 0.06 & 2.8 \\
N4	& 2 54 35.63		& 69 24 43.80		& 0.07 & 2.5 \\
NE1 & 2 54 43.21 		& 69 23 19.78		& 0.08 & 9.5 \\
S1	& 2 54 25.80		& 69 19 47.79		& 0.14 & 34.6 \\
S2	& 2 54 17.49		& 69 19 27.76		& 0.06 & 5.2 \\
S3	& 2 54 25.05		& 69 18 43.79		& 0.09 & 8.6 \\
S4	& 2 54 30.34		& 69 17 47.80		& 0.05 &  2.5 \\
S5	& 2 54 24.31		& 69 17 23.79		& 0.05 &  2.6 \\
SE1	& 2 54 37.89		& 69 19 15.80		& 0.12 & 13.3 \\
SE2	& 2 54 47.70		& 69 19 15.76		& 0.05 & 2.0 \\
SE3	& 2 54 46.94		& 69 18 27.76		& 0.07 &  3.7 \\
\hline \\	
\end{tabular}
\label{core_prop}

\end{center}
\end{table}
%%%%%%%%%%%%%%%%%%%%%%%%%%%%%%%%%%%%%%%%%%%%%%%%%%%%%%%%

\subsection{Magnetic field orientation} \label{magnetic_field}
{\it Planck} polarization data is used to determine the magnetic field orientation in the vicinity of {\g133}. Following the IAU convention, the linear polarization angle (PA) is given by, $\psi_{\rm IAU} = 0.5\times arctan(-U/Q)$ in the Galactic coordinate system \citep{2015A&A...576A.104P}. The plane of sky orientation of the magnetic field is obtained by adding $90\degree$ to the PA ($\chi^\prime=\psi_{\rm IAU}+90\degree$). The position angle of the magnetic field in the equatorial coordinate system, FK5, is then estimated following the discussion in \citet{1998MNRAS.297..617C}, where

\begin{equation}
\psi = arctan~\Bigg[\frac{cos\,\big(l-32.9\degree\big)}{cos\,b~cot\,62.9\degree~-~sin\,b~sin\,\big(l-32.9\degree\big)}\Bigg]
\end{equation}

\noindent
Here, $\psi$ is the angle subtended by the directions of the equatorial north and the Galactic north at each pixel of the {\it Planck} polarization maps. The orientation of the magnetic field at each pixel with Galactic coordinates, $l$ and $b$ is then transformed from the Galactic coordinate system ($\chi^\prime$) to the equatorial system ($\chi$) using the relation,

\begin{equation}
\chi=\chi^\prime-\psi
\end{equation}

\noindent
The mean magnetic field orientation is determined by taking the average of the $\chi$ values that lie within the $3\sigma$ level of the $\rm ^{13}CO\,(1-0)$ emission over the velocity range $-19.4$ to $\rm -11.6\,km\,s^{-1}$. The orientation is estimated to be $99.1 \pm 5.6 \degree$ east of north and is traced in {\fig}\ref{850_clump} with respect to the filaments identified from the 850\,{\um} map. 

\subsection{Associated young stellar population} \label{YSO_population}
\begin{figure}
\centering 
\includegraphics[scale=0.50]{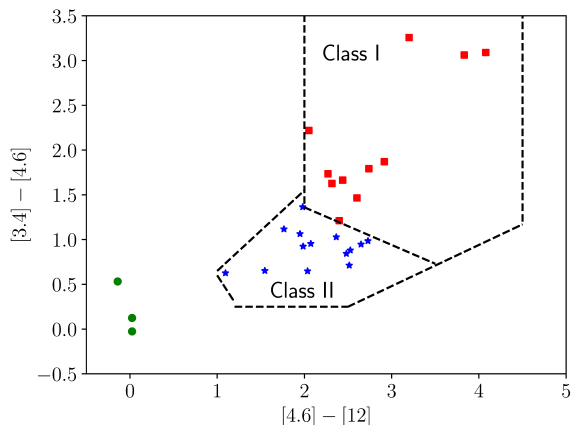}
\caption{WISE color-color diagram for the bands, 3.4, 4.6 and 12\,{\um} used to identify and classify YSOs associated with {\g133}. The dashed lines denotes the criteria used by \citet{2014ApJ...791..131K} to delineate the YSO classes. Green circles denote the sources that do not satisfy the YSO criteria.}
\label{wise_color}
\end{figure}

Using 2MASS-$K_{\rm s}$ and WISE photometry data, \citet{2012MNRAS.421.1040M} have discovered a crowded cluster of 53 YSOs associated with {\g133}. We revisit the cluster detection and YSO identification by considering a larger (radius of 10\,arcmin) region, centred at 02:54:31.4 +69:20:32.5, which effectively samples the entire {\g133} cloud complex. The circular region over which the YSOs are identified is sketched in {\fig}\ref{CO_mom_spec}b and {\fig}\ref{YSO_dist}. For YSO identification, we implement the colour classification scheme of \citet{2014ApJ...791..131K} using the improved ALLWISE catalogue \citep{2013yCat.2328....0C}. After removal of fake or spurious sources and extragalactic contaminants, a WISE colour-colour diagram is constructed and shown in {\fig}\ref{wise_color}. Using the colour criteria set by \citet{2014ApJ...791..131K}, we identify 11 (39\%) Class~I and 14 (50\%) Class~II sources in our field. Sources not satisfying the prescribed criteria could be either Class~III/weak disk YSOs, blue transition disk objects, or AGB stars. Comparing our results with that of \citet{2012MNRAS.421.1040M}, we find that 41 ($\sim 77\%$) of the sources classified as YSOs by \citet{2012MNRAS.421.1040M}, are filtered out as spurious sources on using the stringent and improved source reliability approach proposed by \citet{2014ApJ...791..131K}.
%%%%%%%%%%% color-color %%%%%%%%%%%%
%%%%%%%%%%%%%%%%%%%%%%%%%%%%%%%%

\section{Discussion}
\subsection{Signatures of cloud-cloud collision} \label{signatures}
\par The observed CO morphology and kinematics suggest a cloud-cloud collision scenario for {\g133} with two distinct cloud components, G133a and G133b. We analyse this complex along the lines discussed in \citet{2017ApJ...835L..14G} for the dark cloud L1188. Before we discuss the distinct observational features seen, it is crucial to examine the possibility of the two cloud components being gravitationally bound. The virial mass of the cloud complex can be estimated using the expression \citep{2011A&A...530A.118P},

\begin{equation}
M_{\rm vir} = \frac{5\,\sigma^2\,R}{G}
\end{equation}

\noindent where, $R$ is the effective radius of the cloud and $\rm \sigma (=\Delta V/\sqrt{8\,ln2}) $ is the velocity dispersion. Both these parameters are obtained using the $\rm ^{13}CO\,(1-0)$ line. $R$ is taken to be equal to $(A/\pi)^{0.5}$, where $A$ is the area considered to derive the total cloud mass and is estimated to be $\rm \sim 2.1\,pc$. Fitting a Gaussian profile to the $\rm ^{13}CO\,(1-0)$ spectrum of the entire cloud, we get the line width, $\Delta V$ to be $\rm 3.5\,km\,s^{-1}$. Similar line width is obtained for the spectrum of the intersection region. Taking this, the velocity dispersion is calculated to be $\rm 1.5\,km\,s^{-1}$.
Using these values, we compute the virial mass to be $5.8 \times 10^3\,M_\odot$, which is more than twice the total mass of the cloud complex, $2.6 \times 10^3\,M_\odot$. Within the uncertainties involved in the mass estimations, it is seen that the total mass of the {\g133} cloud complex is not large enough to be able to gravitationally bind the two clouds. This suggests that the physical association of G133a and G133b with each other must be an accidental event.

\par In recent years, numerical hydrodynamical simulations and observational studies \citep[e.g.][]{{2014ApJ...792...63T},{2017ApJ...835..142T},{2018ApJ...859..166F}} have shed light on the various characteristic signatures of cloud-cloud collision. As is evident in {\fig}\ref{CO_mom_spec}, the two clouds, G133a and G133b, are separated by a velocity difference of $\rm \sim 2.8\,km\,s^{-1}$. This separation provides a lower limit to the relative collision velocity. The actual collision velocity might be higher than this because of the projection effect \citep{2015ApJ...807L...4F}. As propounded by \citet{2013ApJ...774L..31I} and \citet{2015ApJ...807L...4F}, the isotropic turbulence is enhanced at the collision-shocked layer, irrespective of the direction of collision. Hence, the velocity spread at the shocked layer can be taken to be similar as the relative collision velocity. Following this, we assume the relative velocity to be the FWHM of the $\rm ^{12}CO\,(1-0)$ line extracted over the intermediate velocity range ($-16.9$ to $-14.1\,\rm km\,s^{-1}$) that corresponds to the shock-compressed layer. The FWHM of this line is found to be $\rm 5.0\,km\,s^{-1}$. Comparing with the velocity difference between the two clouds, the relative collision velocity translates to a relative motion of the two clouds of $\sim 56\degree$ with respect to the line-of-sight. 

\par Another pronounced observational signature of cloud-cloud collision is the ``bridging features" detected in the PV diagram. \citet{{2015MNRAS.450...10H},{2015MNRAS.454.1634H}} made synthetic $\rm ^{12}CO$ PV diagrams using the cloud-cloud collision model data simulated by \citet{2014ApJ...792...63T}. From these observations it was found that the shocked layer is characterised by broad intermediate velocity features, that bridge the colliding clouds in the velocity space. The bridging features often appear at the spots of collision. Several recent studies have provided observational confirmation of such features in the colliding cloud complexes \citep[e.g.][]{{2016ApJ...820...26F},{2017ApJ...835..142T},{2017ApJ...835L..14G}}.
On examining the PV diagram of {\g133} along the two identified directions A and B as depicted in {\fig}\ref{pv_diagram}, we find at least four bridging features clearly visible in the $\rm ^{12}CO\,(1-0)$ and $\rm ^{13}CO\,(1-0)$ emission in the intermediate velocity range of $-16.9$ to $\rm -14.1\,km\,s^{-1}$. The green circles in {\fig}\ref{pv_diagram}(a) indicate the positions where the bridges appear along the PV cuts. These positions correspond to the spots of collision of the two clouds consistent with the discussions in \citet{2018ApJ...859..166F}. The presence of these bridging features suggests the existence of turbulent gas in the shocked layer that is excited by the collision of the clouds. 
In addition, the velocity structure in {\g133} further reveals three high-velocity $\rm ^{12}CO\,(1-0)$ wings, W1, W2 and W3. The locations of these identified wings along the PV cuts are shown in {\fig}\ref{pv_diagram}(a). As discussed by \citet{{2017ApJ...835L..14G},{2019A&A...632A.115G}}, these high-velocity wings could stem from the outflow(s) driven by one or more YSOs or protostellar cores that have been identified in the intersection region (see Section~\ref{core_cluster}). 

\par We examine the cloud at $\rm -10\,km\,s^{-1}$ in the plotted PV diagrams. This cloud displays a rather diffuse morphology in the PV diagrams and is well separated from G133a and G133b. Along the cut B shown in {\fig}\ref{pv_diagram}c, an apparent connecting feature is seen but does not resemble a typical `bridging' structure expected in collision events. Thus, in all likelihood this cloud could well be in the foreground and not physically associated with the {\g133} complex.

\par The morphology of the {\g133} cloud complex as seen in Figs.~\ref{CO_mom_spec}(a)-(c) and \ref{pv_diagram}(a), unveils the picture of an elongated and smaller cloud, G133a, likely in collision with the larger cloud, G133b which displays a `boomerang-like' structure with a cavity opening in the north-east direction. This is in concordance with simulations done by \citet{2010MNRAS.405.1431A}, \citet{1992PASJ...44..203H}, \citet{2014ApJ...792...63T} and \citet{2017ApJ...835..142T}. These authors show that when two clouds of different sizes collide, the smaller cloud creates a cavity on the surface of the larger cloud. Evidence supporting this picture was first perceived in the Galactic H~{\small II} region, RCW\,120 \citep{2015ApJ...806....7T}. Here, the observed infrared bubble (or ring) morphology is well explained by cavity creation due to cloud-cloud collision rather than attributing the same to the conventional expansion of HII region scenario for the bubble origin.

\subsection{Induced filament, core and cluster formation} \label{core_cluster}

%%%%%%%%%%% mass-radius %%%%%%%%%%%%
\begin{figure}
\centering 
\includegraphics[scale=0.53]{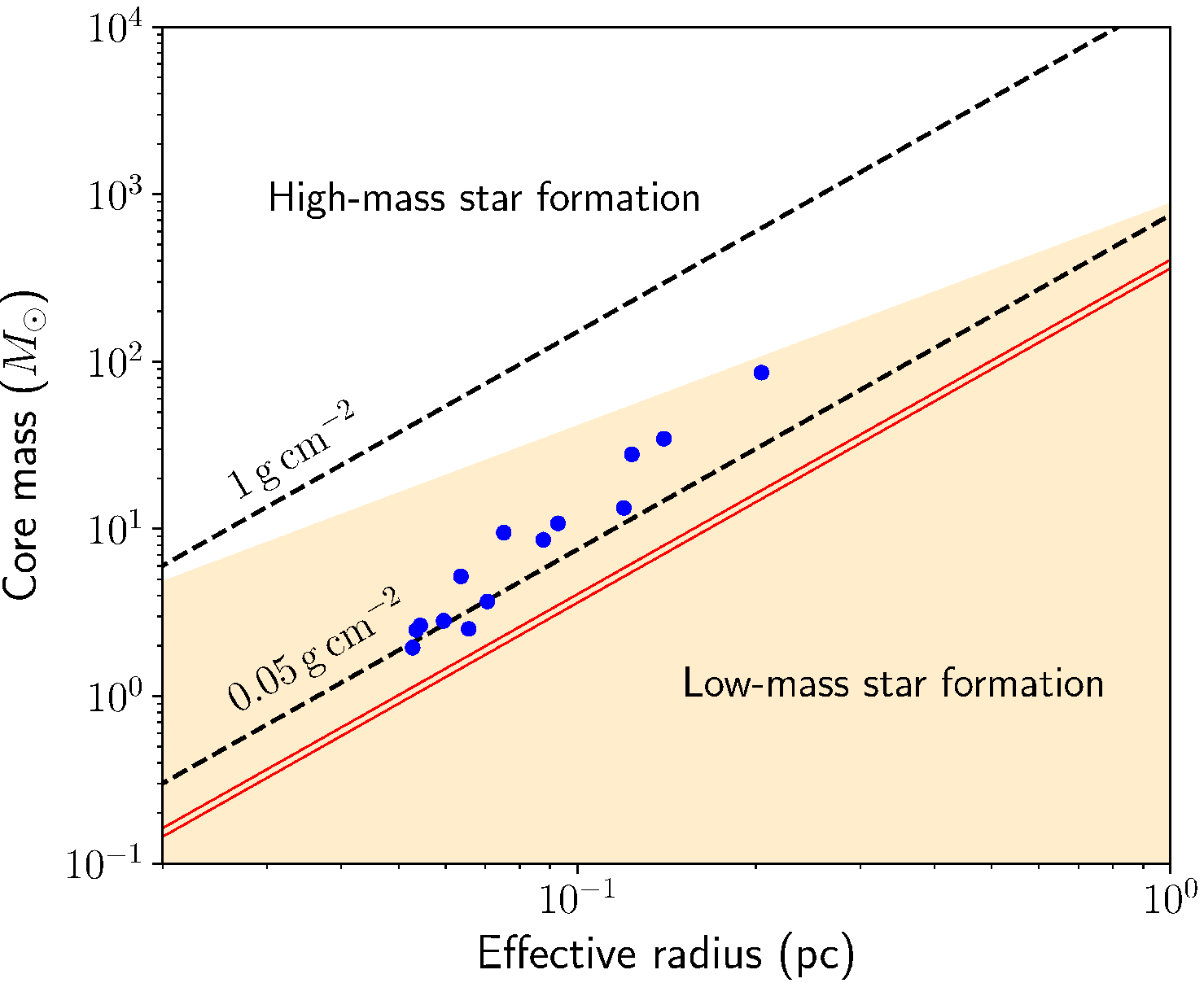}
\caption{The mass of the dense cores, $M_C$, identified from the 850\,{\um} map of {\g133} is plotted as a function of effective radii, $r$ and depicted by blue circles. The red lines indicate the surface density thresholds of $116\,{M_\odot}\,{\rm pc^{-2}} (\rm \sim 0.024\,g\,cm^{-2})$ and $129\,{M_\odot}\,{\rm pc^{-2}} (\rm \sim 0.027\,g\,cm^{-2})$ for active star formation from \citet{2010ApJ...724..687L} and \citet{2010ApJ...723.1019H}, respectively. The shaded region delineates the low-mass star forming region that do not satisfy the criterion $ m(r) > 870\,M_\odot(r/\rm pc)^{1.33} $ \citep{2010ApJ...716..433K}. The black dashed lines represent the surface density threshold of 0.05 and $\rm 1\,g\,cm^{-2}$ defined by \citet{2014MNRAS.443.1555U} and \citet{2008Natur.451.1082K}, respectively.}
\label{mass_radius_cores}
\end{figure}
%%%%%%%%%%%%%%%%%%%%%%%%%%%%%%%%

%%%%%%%%%%% CD-YSO %%%%%%%%%%%%
\begin{figure*}
\centering 
\includegraphics[scale=0.25]{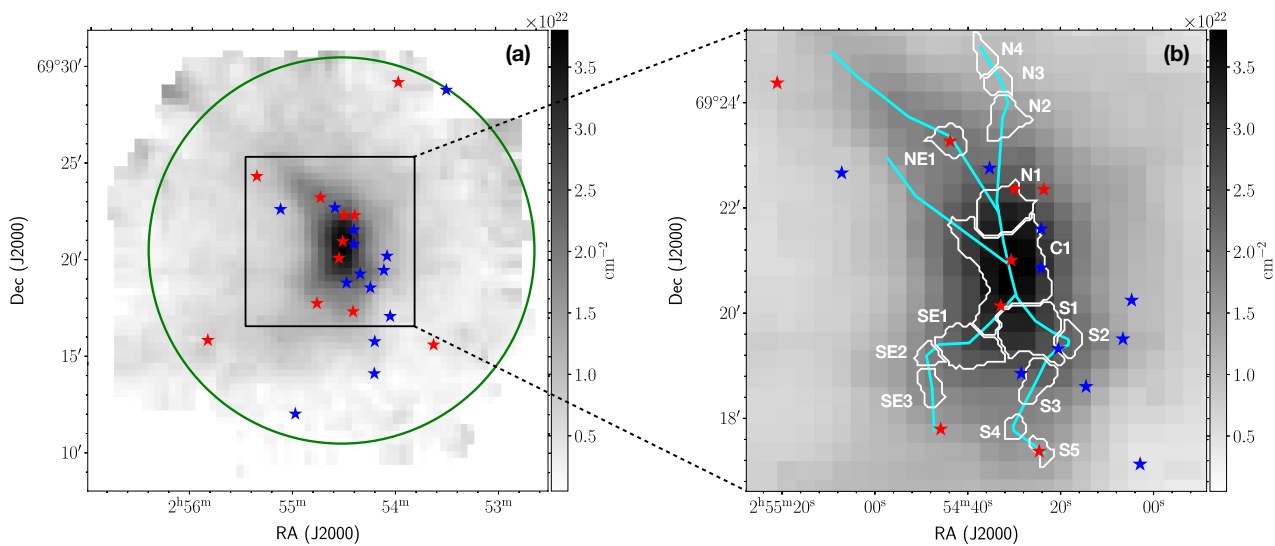}
\caption{(a) The {\h2} column density map same as in {\fig}\ref{Tex_NH2}(b). The green circle indicates the region over which the YSOs are identified. (b) A zoom-in on the arc which has an enhanced column density. The Class~I and Class~II YSOs identified, as discussed in Section~\ref{YSO_population}, are marked by red and blue stars, respectively. The apertures of the 850\,{\um} dense cores and the skeletons of the filaments are overlaid on (b).}
\label{YSO_dist}
\end{figure*}
%%%%%%%%%%%%%%%%%%%%%%%%%%%%%%%%

As discussed in Section~\ref{dust}, distinct filaments are identified which are located in the intersection region of clouds G133a and G133b. As seen in Figs.~\ref{850_clump} and \ref{YSO_dist}(b), the filaments detected in {\g133} present a hub-filament structure. Filament formation is generally understood from the hierarchical collapse of molecular clouds \citep[][and references therein]{2018ARA&A..56...41M}. Following the Hydrodynamic simulations by \citet{1992PASJ...44..203H} and \citet{2010MNRAS.405.1431A}, it is seen that the sites of cloud-cloud collision are characterised by a shock compressed layer due to a bow-shock driven by the smaller cloud into the larger cloud. This augurs well with alternate hypothesis for filament formation that has been presented by \citet{2013ApJ...774L..31I} and \citet{2018PASJ...70S..53I}. Based on their numerical studies using isothermal MHD simulations, these authors conclude that the collision of inhomogeneous clouds lead to the formation of dense filaments in the shock compressed layer, where the magnetic field is amplified in the direction perpendicular to the filament. The clouds are compressed multi-dimensionally except in the direction perpendicular to the background magnetic field, which results in filamentary structures. 

\par In the case of {\g133}, the shock compressed layer created at the interacting front manifests as an open arc structure in the $\rm ^{12}CO\,(1-0)$ and $\rm ^{13}CO\,(1-0)$ maps in the intermediate velocity range, $-16.9$ to $\rm -14.1\,km\,s^{-1}$ as is seen in {\fig}\ref{CO_mom_spec}(b). This arc has a high excitation temperature ($\rm 13-20\,K$), compared to the ambient medium and an enhanced {\h2} column density, evident from {\fig}\ref{Tex_NH2}. The increase in the {\h2} column density within the arc can be due to the multi-dimensional compression of the shocked layer. The filaments detected in the 850\,{\um} map of {\g133} follows the same orientation of the arc structure. The direction of the magnetic field in the region associated with {\g133} is determined in Section~\ref{magnetic_field}, and is sketched in {\fig}\ref{850_clump}. Concurrent with the results from the MHD simulation by \citet{2013ApJ...774L..31I}, the background magnetic field, probed using {\it Planck} data, is seen to be oriented perpendicular to the filamentary structure. It should be noted that the arc could also be the result of stellar feedback (radiation pressure, expanding HII region, outflows from YSOs). From our literature survey, we find no HII region associated with {\g133}, however given the YSOs detected, one cannot conclusively rule out stellar feedback. 

\par Collision induced enhanced density in the intersection region triggers formation of dense cores that accrete matter from the natal filaments. We have detected 14 dust cores in {\g133}, the masses of which are estimated and found to lie within $\rm \sim 2 - 86\,M_\odot$ with sizes ranging between $\rm \sim 0.05 - 0.2\,pc$. In comparison, \citet{2018ApJS..236...49Z} detect 18 dust cores from the 850\,{\um} continuum map. They use the GaussClumps algorithm for core identification, where only the cores with peak intensities above $5\sigma$ are considered. Correlating the dust cores identified by these authors with the ones we extracted using the FellWalker algorithm, we find that majority of the core peaks match. Further, the central core, C1, detected by us encompasses five core peaks identified by \citet{2018ApJS..236...49Z}.

In {\fig}\ref{mass_radius_cores}, which is adapted from \citet{Yuan2017} and \citet{2020MNRAS.496.2790L}, we correlate the mass and effective radius of the clumps and interpret their nature.  As is discernible from the figure, the dust cores associated with {\g133} lie comfortably above the mass surface density threshold for active star formation proposed by \citet{2010ApJ...724..687L} and \citet{2010ApJ...723.1019H}. Probing the mass regime possible, we examine the empirical relation, $ m(r) > 870{\rm M_\odot}(r/\rm pc)^{1.33}$, given by \citet{2010ApJ...716..433K} for high-mass star forming cores. From the estimated mass and radius, the cores associated with {\g133} do not qualify this threshold and hence should be devoid of high-mass stars. This is somewhat contrary with regards to the limit of  0.05 $\rm g\,cm^{-2}$ on surface density set by \citet{2014MNRAS.443.1555U} for massive star formation. Barring one, all the active star-forming cores are above this threshold. Hence, it is likely that the more massive cores in our sample are potential high-mass star-forming regions.  

\par We also probe the star-formation activity associated with {\g133} by inspecting the distribution of the identified YSOs in the region. In {\fig}\ref{YSO_dist}(a), we show the YSOs overlaid on the column density map. Of the 25 YSOs detected, a distinct cluster of 18 YSOs is seen to be located in the intersection region.  {\fig}\ref{YSO_dist}(b) shows an enlarged view with the detected filaments and cores overlaid. From the relative collision velocity of the clouds which is estimated to be $\rm 5.0\,km\,s^{-1}$, we compute the timescale of collision between G133a and G133b. The ratio of the cloud size, 4.2\,pc, to the relative collision velocity gives a collision timescale of $\rm \sim 0.8\,Myr$. It should however be noted that this gives an order-of-magnitude estimate, at best. The value might vary by a factor of $\sim 2$ owing to projection effects in space and velocity and to the unknown configuration of the clouds before collision \citep{2014ApJ...780...36F}. Nonetheless, the timescale derived is longer than the typical lifetimes of Class~I YSOs and is comparable to the lifetime of Class~II YSOs. As discussed in \citet{2015ApJS..220...11D} and \citet{2009ApJS..181..321E}, the age estimates of Class I and Class II YSOs are 0.4-0.7\,Myr and $\rm 2\pm 1\,Myr$, respectively. The scenario that has unfolded in {\g133} gives compelling evidence of this being a bona fide case of collision induced cluster formation. Similar results are obtained in several other studies focused towards cloud collision and induced cluster formation \citep[e.g.][]{{2011ApJ...738...46T},{2017ApJ...835L..14G}}. 

\section{Conclusions}

%%%%%%%%%%% schematic %%%%%%%%%%%%
\begin{figure*}
\centering 
\includegraphics[scale=0.17]{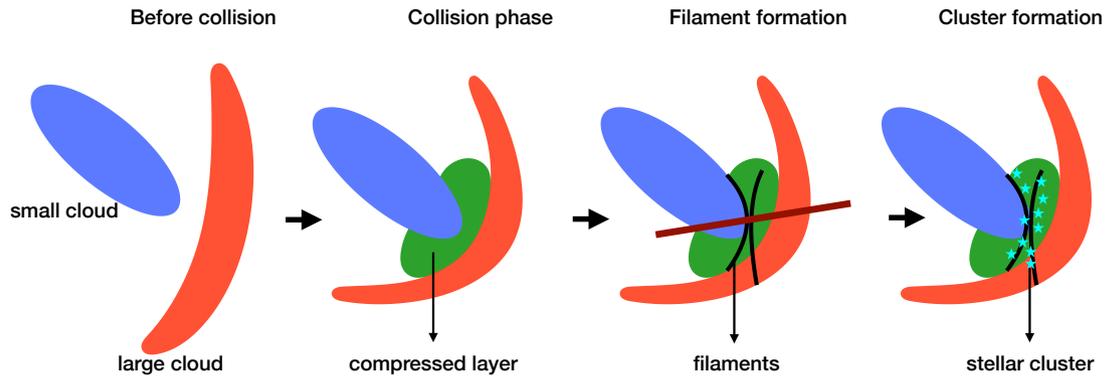}
\caption{A schematic of the cloud-cloud collision in {\g133} depicting the scheme of events from the time of collision of G133a (small cloud) with G133b (large cloud). The orientation of the background magnetic field is indicated by the maroon line.}
\label{schematic}
\end{figure*}
%%%%%%%%%%%%%%%%%%%%%%%%%%%%%%%%

Gas kinematics studied using the CO lines show a picture of colliding clouds G133a and G133b triggering the formation of a complex network of filaments, dense cores and YSOs in {\g133} complex. {\fig}\ref{schematic} depicts a schematic of the chain of events occurring. Clouds G133a and G133b with a velocity difference of $\rm \sim 2.8\,km\,s^{-1}$ are likely to have collided forming a shock-compressed layer in the intersection region. Conforming to MHD simulations, the formation of a complex network of filaments is deduced from the 850\,{\um} map. 14 dust cores that accrete matter from the natal filaments are also identified. The over-density of Class I and II YSOs along the intersection arc advocates for collision induced cluster formation in {\g133}. Keeping in mind that the PMO observations presented in this paper might not have sampled the entire complex, an accurate interpretation of the morphology of the colliding clouds is difficult. Further, the resolution of the data used is not adequate to correlate the observed scenario with either the spherical cloud collision models \citep[e.g.][]{{1992PASJ...44..203H},{2014ApJ...792...63T}} or filamentary cloud collision simulations \citep{2018MNRAS.473.4220L}. Nonetheless, observational features seen makes {\g133} an interesting candidate to probe cloud-cloud collision. With follow-up high-resolution molecular line and dust continuum observations, viable models can be explored to explain the observed signatures.
\section*{Acknowledgements}

We would like to thank the referee for valuable comments/suggestions. The authors would like to thank Dr. Pak Shing Li for fruitful discussions. We are grateful to the staff at the Qinghai Station of PMO for their assistance during the observations. The James Clerk Maxwell Telescope is operated by the East Asian Observatory on behalf of The National Astronomical Observatory of Japan; Academia Sinica Institute of Astronomy and Astrophysics; the Korea Astronomy and Space Science Institute; Center for Astronomical Mega-Science (as well as the National Key R\&D Program of China with No. 2017YFA0402700). Additional funding support is provided by the Science and Technology Facilities Council of the United Kingdom and participating universities in the United Kingdom and Canada. Additional funds for the construction of SCUBA-2 were provided by the Canada Foundation for Innovation. This publication makes use of data products from the {\it Wide-field Infrared Survey Explorer}, which is a joint project of the University of California, Los Angeles, and the Jet Propulsion Laboratory/California Institute of Technology, funded by the National Aeronautics and Space Administration.

\section*{DATA AVAILABILITY}

The original data underlying this article will be shared on reasonable request to the corresponding author.
%%%%%%%%%\g%%%%% %%%%%%%%%%%%%%%%%%%%%%%%%%%%%%%%%%%%



\bibliographystyle{mnras}
\bibliography{reference} % if your bibtex file is called reference.bib

%%%%%%%%%%%%%%%%%%%%%%%%%%%%%%%%%%%%%%%%%%%%%%%%%%

%%%%%%%%%%%%%%%%% APPENDICES %%%%%%%%%%%%%%%%%%%%%

%\appendix


%%%%%%%%%%%%%%%%%%%%%%%%%%%%%%%%%%%%%%%%%%%%%%%%%%

% Don't change these lines
\bsp	% typesetting comment
\label{lastpage}
\end{document}